\documentclass[aip,jcp,reprint,amsmath,amssymb,floatfix,citeautoscript]{revtex4-1}
\usepackage{amsmath}
\usepackage{graphicx}
\usepackage{amssymb}
\usepackage{amsthm}
\usepackage{bm}

\begin{document}

\title{Reduced density matrix hybrid approach: Application to electronic energy transfer}

\author{Timothy C. Berkelbach}
\email{tcb2112@columbia.edu}
\affiliation{Department of Chemistry, Columbia University, 3000 Broadway, New York, New York 10027, USA}

\author{Thomas E. Markland}
\affiliation{Department of Chemistry, Stanford University, 333 Campus Drive, Stanford, California 94305, USA} 

\author{David R. Reichman}
\affiliation{Department of Chemistry, Columbia University, 3000 Broadway, New York, New York 10027, USA}

\begin{abstract}
Electronic energy transfer in the condensed phase, such as that occurring in photosynthetic complexes, frequently
occurs in regimes where the energy scales of the system and environment are similar. This situation provides a challenge to
theoretical investigation since most approaches are accurate only when a certain energetic parameter is small
compared to others in the problem. Here we show that in these difficult regimes, the Ehrenfest approach provides
a good starting point for a dynamical description of the energy transfer process due to its ability to
accurately treat coupling to slow environmental modes. To further improve on the accuracy of the Ehrenfest approach,
we use our reduced density matrix hybrid framework to treat the faster environmental modes quantum mechanically,
at the level of a perturbative master equation. This combined approach is shown to provide an efficient
and quantitative description of electronic energy transfer in a model dimer and the Fenna-Matthews-Olson
complex and is used to investigate the effect of environmental preparation on the resulting dynamics.
\end{abstract}

\maketitle

\section{Introduction}\label{sec:intro}
Recent experimental observations of long-lived electronic coherence in photosynthetic complexes\cite{eng07,lee07}
and solutions of conjugated polymers\cite{col09_1,col09_2} have challenged the
conventional view that environmental effects rapidly quench quantum coherence at room temperature.
These experiments have spurred the theoretical investigation of electronic energy transfer (EET), including
the development of accurate numerical methodologies. Such work has sought to explain the origin of
the observed coherence lifetime, to predict the effects of system parameters, and to ultimately
understand the role of quantum coherence in promoting or inhibiting efficient energy
transfer\cite{che09,ish10,yan02,moh08,reb09,ish09_jcp1,ish09_jcp2,ish09_pnas,jan08,huo10,tao10,jan11,
kel11,ren11,nal11,olb11_jpcl,olb11_jpcb}.

EET systems typically consist of a set of molecular chromophores which are electronically coupled to
one another as well as to local environmental phonons. The former couplings facilitate exciton delocalization
while the latter couplings tend to destroy this so-called quantum coherence. Hence, accurately modeling the interplay 
between these effects is crucial.  A difficulty which arises in the modeling of EET is the similarity of energy
scales describing these competing phenomena.  For example, in typical multi-chromophoric systems,
the electronic couplings, reorganization energies, and characteristic environmental frequencies
are all on the order of $10-100\ {\rm cm}^{-1}$.

As examples of perturbative methods which may fail in this regime,
the popular Redfield\cite{red65,ish09_jcp1} and F\"{o}rster\cite{for53} theories of energy transfer implicitly
require nearly Markovian, non-adiabatic dynamics characterized by environmental frequencies much larger than either
the reorganization energy or electronic couplings, respectively.  It is one of the goals of this work to show that
the violation of this expectation in the form of atypically small environmental frequencies actually suggests
a promising route towards accurate theories of EET in this intermediate coupling regime.
Specifically, quantum-classical theories are well suited to the problem of EET,
where environmental fluctuations take place over timescales on the order of the electronic motion.
In such approaches, the environmental degrees of freedom are treated classically since quantum effects are
expected to be insignificant for such low-frequency vibrational motion.  Most relevant to EET,
these methods can accurately describe non-Markovian effects such as transport mediated by non-equilibrium phonon states.
Investigations along this line have included the application of LSC-IVR\cite{tao10}, the Poisson
bracket mapping formalism\cite{kel11}, iterative linearized propagation schemes\cite{huo10}, and a modified
variant of mean-field Ehrenfest dynamics\cite{ish11}. In a similar vein, the recently developed reduced
hierarchy equations\cite{ish09_jcp2}, although quantum-mechanically exact when fully converged, are
numerically simplest for systems at high-temperature or with a slow, adiabatic bath.

In accord with the above discussion, in the present work we show that a simple quantum-classical
Ehrenfest treatment of EET systems yields results in qualitative and even sometimes semi-quantitative
agreement with existing exact results. In particular, quantitative agreement is found for short-time
dynamics, as well as coherence frequencies and lifetimes. This latter finding provides insight into the mystery
of long-lived quantum-coherence: the intramolecular motions are simply too slow to induce effective dephasing.
We also demonstrate that including the quantum subsystem's back-reaction on the classical environment
improves the long-time populations in contrast compared to a previous Ehrenfest study where this effect
was not included \cite{ish11}. However, although this improves the accuracy of long-time populations,
the Ehrenfest method generally still yields incorrect long-time values due to a well-known intrinsic violation
of detailed balance.
These long-time properties are crucial for the accurate description of
EET, where the population of target sites provides a simple metric for the overall transport efficiency.

To address this issue, we invoke our recently introduced reduced density matrix hybrid (RDM-Hybrid)
methodology\cite{ber11_1} which enables one to treat the system and high frequency environmental modes
quantum mechanically, while the slow modes are handled by the Ehrenfest approach. This methodology
allows for efficient treatment of {\em quantum and classical} environmental modes and is able to
quantitatively correct the discrepancies of Ehrenfest dynamics, yielding excellent agreement
with the exact results obtained by Ishizaki and Fleming\cite{ish09_jcp2,ish09_pnas} using the reduced
hierarchy equations (RHE).  As an example of the types of problems one may further probe using our
physically transparent methodology, we investigate the effects of the initial bath preparation on
the subsequent observation of coherent quantum dynamics.

The outline of the paper is as follows.  In Sec.~\ref{sec:ham} we present the Frenkel exciton Hamiltonian
used for EET modeling.  We then review the Ehrenfest and RDM-Hybrid methodologies in Sec.~\ref{sec:review}.
In Sec.~\ref{sec:niba} we derive a perturbative quantum master equation to use in our RDM-Hybrid algorithm
for the system and quantum environment modes.  We present Ehrenfest and RDM-Hybrid results in Sec.~\ref{sec:results},
including population dynamics and rate constants for a simple dimer, as well as population dynamics for the
seven-site Fenna-Matthews-Olson complex with two types of environmental preparation.  We conclude in Sec.~\ref{sec:conc}.

\section{Model Hamiltonian}\label{sec:ham}

As in other theoretical work on EET, we adopt the Frenkel exciton Hamiltonian for $N$ chromophores, given by
\begin{equation}\label{eq:ham1}
H = \sum_{n=1}^N |n\rangle E_n \langle n | + \sum_{m\neq n}^N |m\rangle J_{mn}\langle n| + H_b + H_{sb}
\end{equation}
where one assumes independent baths for each site, 
\begin{equation}\label{eq:ham2}
H_b = \sum_{n=1}^N \sum_{k} \left[ P_{k,n}^2 + \omega_{k,n}^2 Q_{k,n}^2 \right]/2,
\end{equation}
and a bi-linear system-bath coupling,
\begin{equation}\label{eq:ham3}
H_{sb} = \sum_{n=1}^N |n\rangle\langle n| \sum_{k} c_{k,n} Q_{k,n}.
\end{equation}
Such a Hamiltonian physically describes the single-excitation subspace of the complex's total Hilbert space, where
$E_n$ denotes the energy of the total system when the $n$th chromophore is excited and all others are in their ground state.
The electronic coupling of this excitation between sites $m$ and $n$ is given by $J_{mn}$ and is assumed to be static.

In contrast to the spin-boson Hamiltonian investigated in our previous work, which described a two-level system coupled
to one shared bath\cite{ber11_1}, the present Hamiltonian is more commonly adopted for molecular energy transfer,
where molecular vibrations
and surrounding environmental effects are local and approximately uncorrelated.

The phonon bath located at each site is completely characterized by its spectral density, which we take to be of the
Debye form,
\begin{align}
J_n(\omega) &= \frac{\pi}{2} \sum_{k} \frac{c_{k,n}^2}{\omega_{k,n}} \delta(\omega-\omega_{k,n}) \nonumber\\
	& = 2 \lambda \omega_c \frac{\omega}{\omega^2 + \omega_c^2}. \label{eq:debye}
\end{align}
In the above, $\omega_c$ is the bath cutoff frequency and the
reorganization energy, $\lambda = \pi^{-1} \int_0^\infty d\omega J(\omega)/\omega$,
is the energy dissipated by the environment after a Franck-Condon transition of the excitation from one site to another.
In the second line and henceforth, we assume that all sites have the same spectral density. However we note that the
methods employed here are in no way limited (or catered) to the treatment of independent baths, the functional form of the spectral density,
or the assumption of identical baths.  Future work will include the investigation of each of these effects on energy
transfer dynamics.

\section{Review of Ehrenfest and RDM-Hybrid methods}\label{sec:review}
In this section, we briefly review the pertinent details of the Ehrenfest and reduced density matrix hybrid (RDM-Hybrid)
quantum dynamics methodologies needed to treat to the Hamiltonian in Eqs.~(\ref{eq:ham1})-(\ref{eq:ham3}). For full details of the
method, we refer the reader to our previous paper\cite{ber11_1}.

\subsection{Ehrenfest method}
In the Ehrenfest method, one assumes separability of system and bath variables, yielding the product density matrix
$\rho(t) \approx \rho_s(t) \rho_b(t)$.  Inserting this ansatz into the Liouville
equation and tracing out the system or bath variables yields the time-dependent self-consistent
field coupled equations of motion.  A classical treatment of the bath density operator then yields the Ehrenfest
equations of motion,
\begin{align}
\frac{\partial \rho_s(t)}{\partial t} &= -i\left[ H_s,\rho_s(t) \right] \nonumber\\
	&\hspace{1em} -i\left[ \sum_{n=1}^N |n\rangle\langle n| \sum_{k} c_{k,n} Q_{k,n}(t), \rho_s(t) \right], \\
\frac{d Q_{k,n}}{dt} &= P_{k,n} \\
\frac{d P_{k,n}}{dt} &= -\omega_{k,n}^2 Q_{k,n} - c_{k,n} {\rm Tr}_s \big\{ |n\rangle\langle n| \rho_s(t) \big\} \label{eq:batheom2}
\end{align}
where square brackets denote the commutator.
The fluctuating bath coordinates yield a time-dependent bias for the system Hamiltonian, and in-turn the system populations
${\rm Tr}_s \big\{ |n\rangle \langle n| \rho_s(t) \big\} = P_n(t)$ yield a driving force for the bath coordinates.
The system populations $P_n(t)$ are averaged over an ensemble of such coupled trajectories
with the harmonic bath initial conditions sampled from the classical Boltzmann or the quantum Wigner distribution\cite{ber11_1}.

\subsection{RDM-Hybrid method}

The RDM-Hybrid method differs from the Ehrenfest approach by partitioning the bath degrees of freedom into `core' and
`reservoir' modes.  In terms of the spectral density, we have
\begin{align}
J_{\rm core}(\omega) &= J(\omega)\left[1 - S(\omega,\omega^*)\right], \\
J_{\rm res}(\omega) &= J(\omega)S(\omega,\omega^*),
\end{align}
where $S(\omega,\omega^*)$ is a switching function, taken here to be
\begin{equation}\label{eq:smoothswitch}
S(\omega,\omega^*) = \begin{cases}
\left[1 - \left(\omega/\omega^*\right)^2\right]^2 & \omega < \omega^* \\
0 & \omega > \omega^*,
\end{cases}
\end{equation}
which switches smoothly from 1 to 0 as $\omega$ goes from 0 to $\omega^*$.  The switching frequency, $\omega^*$ is taken
to be a characteristic timescale of the electronic system\cite{ber11_1}.

Positing the separation of `system-core' and `reservoir' density operators, $\rho(t) \approx \rho_{sc}(t) \rho_r(t)$,
one finds a Liouville equation for the density matrix of the system and core modes,
$\rho_{sc}(t)$, given by
\begin{equation}
\frac{d\rho_{sc}(t)}{dt} = -i \left[ H_{sc}(t), \rho_{sc}(t) \right],
\end{equation}
where the time-dependent Hamiltonian, $H_{sc}(t)$, is a modified system-core Hamiltonian,
\begin{align}
H_{sc}(t) &= \sum_{n=1}^N |n\rangle E_n(t) \langle n | + \sum_{m\neq n}^N |m\rangle J_{mn}\langle n|	\nonumber\\
	&\hspace{1em} + \sum_{n=1}^N \sum_{k \in {\rm core}} \Bigg\{ \frac{1}{2} \left[ P_{k,n}^2 + \omega_{k,n}^2 Q_{k,n}^2 \right] \nonumber\\
		&\hspace{9em} + |n\rangle\langle n| c_{k,n} Q_{k,n} \Bigg\},
\end{align}
and the time-dependent bias arises from the coupling to the classical reservoir modes,
\begin{equation}
E_n(t) = E_n + \sum_{k \in {\rm res}} c_k Q_k(t).
\end{equation}
A solution of the above Liouville equation for the total system-core density matrix is of course intractable, but its
{\em reduced density matrix} averaged over the core degrees of freedom can be calculated by a variety of existing
approximate and exact methods.  The diagonal elements (populations) of this reduced density matrix,
\begin{equation}
{\rm Tr}_s {\rm Tr}_c \big\{|n\rangle\langle n| \rho_{sc}(t)\big\} = {\rm Tr}_s \big\{ |n\rangle\langle n| \rho_s(t) \big\} = P_n(t),
\end{equation}
in turn yield a driving force in the classical reservoir equations of motion,
\begin{align}
\frac{d Q_{k,n}}{dt} &= P_{k,n} \\
\frac{d P_{k,n}}{dt} &= -\omega_{k,n}^2 Q_{k,n} - c_{k,n} P_n(t).
\end{align}
As in the Ehrenfest method, the final system populations are calculated as an average over trajectories.
In this work, the initial conditions of the reservoir modes are sampled
from the Wigner distribution. To evolve the reduced system-core density matrix we perform an approximate calculation using a 
perturbative quantum master equation. This execution of the RDM-Hybrid approach yields an efficient methodology
not much more expensive than a
typical master equation calculation but with far superior accuracy

As shown in our previous paper\cite{ber11_1}, the RDM-Hybrid
methodology naturally interpolates between the regimes of validity of its composite methods and furthermore works well even in regimes
where neither method alone is accurate.  By employing the Ehrenfest method, accurate for nearly adiabatic dynamics, and a
non-adiabatic quantum master equation, the RDM-Hybrid approach can accurately treat the entirety of parameter space using
a single dynamical scheme.
The perturbative master equation used in this work to treat the 
driven system-core dynamics is derived in the following section. 

\section{Master equation for system and core}\label{sec:niba}

Following the success of our previous work\cite{ber11_1}, which employed the noninteracting blip approximation (NIBA)\cite{wei08,leg87}
for the system-core reduced dynamics, we here derive a multi-site generalization, perturbative to 
second order in the electronic couplings, $J_{mn}$,
which we shall continue to refer to as `NIBA,' for simplicity.  A similar master equation, the 
noninteracting cluster approximation\cite{egg94}, has been derived by different means.

Following Hu and Mukamel\cite{hu89} (see also Golosov and Reichman\cite{gol01}), we define a Liouville-space projection operator,
\begin{equation}
\mathbf{P} = \sum_{n=1}^{N} |n \rho_b\rangle\rangle \langle\langle n|.
\end{equation}
Here $|n\rangle\rangle = |n\rangle1_b\langle n|$, $1_b$ is the identity operator in the bath degrees of freedom,
$|n\rho_b\rangle\rangle = |n\rangle \rho_b \langle n|$, and the inner product is given by
$\langle\langle A | B \rangle \rangle = {\rm Tr}_s {\rm Tr}_b (A^\dagger B)$.  In this notation, the observables of interest,
i.e. the site populations, are written as $P_n(t) = \langle\langle n | \rho(t) \rangle\rangle$, where $\rho(t)$ is the
total system-bath density matrix.  

Employing the usual projection operator formalism\cite{bre02,hu89,gol01}, one may derive the exact set of equations
\begin{equation}\label{eq:niba}
\dot{P}_n(t) = \sum_{m=1}^{N} \int_0^t d\tau K_{nm}(t,\tau) P_m(\tau)
\end{equation}
where the kernels are given by
\begin{equation}
K_{nm}(t,\tau) = \langle\langle n | \mathcal{L}_V(t) U(t,\tau) \mathcal{L}_V(\tau) | m\rho_b \rangle\rangle.
\end{equation}
In the above expression, $\mathcal{L}_V(t)\dots = [V(t),\dots]$ is the Liouvillian in the interaction picture,
i.e. $V(t) = \exp(i H_0 t) V \exp(-i H_0 t)$. The propagator is given by
\begin{equation}
U(t,\tau) = \exp_T \left[ -i \int_\tau^t dt^\prime \mathbf{Q} \mathcal{L}_V(t^\prime) \right],
\end{equation}
with $\exp_T[\dots]$
denoting the usual time-ordered exponential and $\mathbf{Q} = \mathbf{1}-\mathbf{P}$ is the complementary projection operator.
The perturbation here is $V = \sum_{m\neq n}^N |m\rangle J_{mn} \langle n|$ with the unperturbed
Hamiltonian given simply by $H_0 = H-V$.

While the above formalism is exact, the propagator $U(t,\tau)$ in the complementary subspace is intractable.  However
the propagator may be expanded perturbatively, and to lowest non-trivial order in the electronic couplings (obtained by
setting $U(t,\tau)=1$),
one obtains the second-order kernels,
\begin{align}\label{eq:kernel}
K_{n\neq m}(t,\tau) &= 2 J_{nm}^2 \exp\left[-Q_2(t-\tau)\right] \nonumber\\
	&\hspace{-3em} \times \cos\Bigg\{\zeta_n(t,\tau)-\zeta_m(t,\tau) + Q_1(t-\tau) \nonumber\\
	&\hspace{0em} - \left(1+\frac{\delta_n + \delta_m}{2}\right)\left[Q_1(t) - Q_1(\tau)\right]\Bigg\}
\end{align}
with $K_{nn}(t,\tau) = -\sum_{m\neq n}^N K_{mn}(t,\tau)$.
In the above, we have introduced the site-dependent accumulated phase,
\begin{align}
\zeta_n(t,\tau) & = \int_\tau^t dt^\prime E_n(t^\prime) \nonumber\\
	&= \int_\tau^t dt^\prime \left[ E_n + \sum_{k \in {\rm res}} c_{k,n} Q_{k,n}(t^\prime) \right]
\end{align}
and the bath correlation function, $Q(t)=Q_2(t)+iQ_1(t)$, is given by
\begin{align}
Q(t) &= \frac{2}{\pi} \int_0^\infty d\omega \frac{J_{\rm core}(\omega)}{\omega^2}
		\left\{ \coth(\beta\omega/2)\left[1-\cos(\omega t)\right]\right. \nonumber\\
	&\hspace{10em}\left.+\ i\sin(\omega t) \right\}.
\end{align}
The factor of two appearing here differs from the usual
NIBA expressions due to the assumption of independent baths.  Lastly, the shift parameters $\delta_n$ arise
from the initial bath density matrix,
\begin{equation}\label{eq:bathic}
\rho_b = \prod_n Z_n^{-1} \exp \left( - \beta \sum_{k} h_{k,n} \right)
\end{equation}
where
\begin{equation}\label{eq:bathshift}
h_{k,n} = \frac{1}{2}\left[ P_{k,n}^2 + \omega_{k,n}^2 \left(Q_{k,n} - \delta_n c_{k,n} \right)^2 \right]. 
\end{equation}

We now demonstrate that our generalized master equation, Eq.~(\ref{eq:niba}), naturally reduces to the hopping
rate equation predicted by F{\" o}rster theory in its regime of validity. Specifically, we consider
the strongly non-adiabatic regime,
such that none of the bath modes are treated classically, i.e. $\zeta_n(t,\tau) = E_n\times(t-\tau)$.
For simplicity we consider the case $\delta_n = \delta_m = -1$ such that the final term in Eq.~(\ref{eq:kernel}) vanishes.

When the bath dynamics are much faster than those of the electronic subsystem, the memory kernel in
Eq.~(\ref{eq:niba}) decays very quickly, such that the Markovian approximation can be made
\begin{equation}
\dot{P}_n(t) \approx \sum_{m=1}^{N} k_{nm} P_m(t)
\end{equation}
where the rate constant $k_{nm}$ is given by $k_{nm} = \int_0^\infty dt K_{nm}(t)$.  Introducing the
well-known line-shape function $g(t) = Q(t)/2 - i\lambda t$, with $\lambda$ the reorganization energy, we can write
\begin{align}
k_{nm} &= 2 J_{nm}^2 {\rm Re} \int_0^\infty dt e^{-i (E_n +\lambda) t - g(t)} e^{i (E_m -\lambda)t - g(t)} \nonumber\\
	&= J_{nm}^2 \int_{-\infty}^\infty dt \int_{-\infty}^\infty dt^\prime \delta(t-t^\prime) \nonumber\\
	&\hspace{4em} \times e^{-i (E_n + \lambda) t - g(t)} e^{i (E_m - \lambda) t^\prime - g(t^\prime)}.
\end{align}
Using the Fourier transform of the Dirac delta function,
$\delta(t-t^\prime) = (2\pi)^{-1} \int_{-\infty}^{\infty} d\omega \exp[ i \omega (t-t^\prime) ]$,
we have
\begin{equation}\label{eq:forster}
k_{nm} = \frac{J_{nm}^2}{2\pi}\int_{-\infty}^\infty d\omega A_n(\omega) F_m(\omega)
\end{equation}
with the absorption and fluorescence spectra given by
\begin{align}
A_n(\omega) &= \int_{-\infty}^\infty dt e^{i\omega t} e^{-i (E_n + \lambda) t - g(t)}, \\
F_m(\omega) &= \int_{-\infty}^\infty dt e^{i \omega t} e^{-i (E_m - \lambda) t - g^*(t)}.
\end{align}
Note we have made use of the fact that $A(\omega)$ and $F(\omega)$ are real, as can be explicitly checked
by symmetry. Equation~(\ref{eq:forster}) is readily recognized as the celebrated F{\"o}rster rate.

\section{Results}\label{sec:results}

\subsection{Model EET dimer}

We begin our application of the Ehrenfest and RDM-Hybrid methodologies with the simplest model of excitation energy
transfer, a dimer.  For the sake of comparison, we employ the model Hamiltonian, Eqs.~(\ref{eq:ham1})-(\ref{eq:ham3}),
with a Debye spectral density, Eq.~(\ref{eq:debye}), and the parameters of Ishizaki and
Fleming\cite{ish09_jcp2}, for which Redfield theory has already been shown to fail badly\cite{ish09_jcp2}.
In the results to follow, we compare to existing results obtained via the numerically exact reduced hierarchy equations
(RHE)\cite{ish09_jcp2}, based on work originally done by Kubo et al\cite{tak77,tan89}.
For the physically motivated reasons explained in our previous paper\cite{ber11_1}, we take the the
splitting frequency to be equal to the Rabi frequency of the electronic subsystem,
\begin{equation}\label{eq:wstar}
\omega^* = \omega_R = \sqrt{\left(E_1-E_2\right)^2 + 4J_{12}^2}.
\end{equation}

In all of our results we discretized the bath into $f=300$ modes with frequencies and couplings
given by
\begin{equation}
\omega_k = \omega_c \tan\left[ \frac{\pi}{2f} \left( k - 1/2 \right) \right],
\end{equation}
\begin{equation}
c_k^2 = \frac{2}{\pi} \omega_k \frac{J(\omega_k)}{\rho(\omega_k)} = 2 \lambda \omega_k^2 / f,
\end{equation}
for $k=1,2,\dots,f$, which can be shown to reproduce the reorganization energy exactly.  The coupled
system-bath equations of motion were solved with a second-order Runge-Kutta scheme using a timestep of 0.5 fs.
For consistency, the classical degrees of freedom were sampled from the Wigner distribution, though for the
relatively high temperatures considered here, the results are largely insensitive to this choice
when compared to those obtained by purely classical Boltzmann sampling.

\begin{figure}[t]
\centering
\includegraphics[scale=0.4]{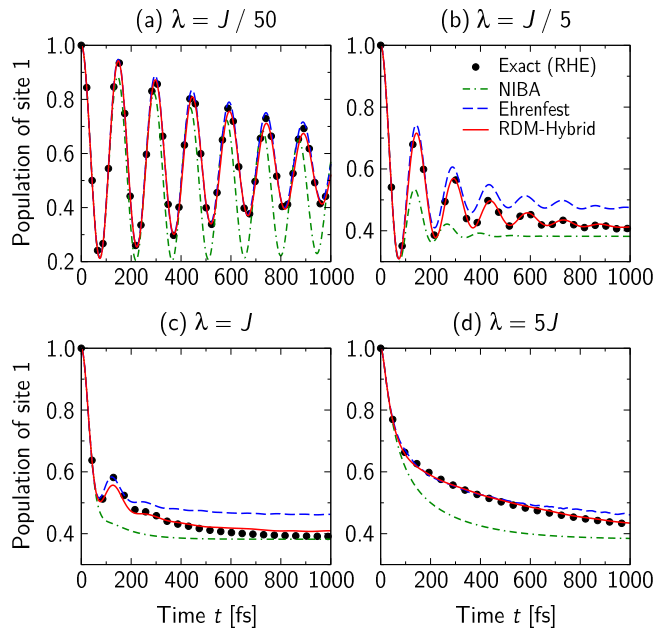}
\caption{Population of site 1 in an EET dimer system considered by Ishizaki and Fleming
with $E_1-E_2 = 100\ {\rm cm}^{-1}$, $J = 100\ {\rm cm}^{-1}$, $\omega_c = 53\ {\rm cm}^{-1}$
($\omega_c^{-1} = 100$~fs), and $T = 300\ {\rm K}$.
Each site is coupled to its own bath with a Debye spectral density.
}
\label{fig:ishizaki4}
\end{figure}

\subsubsection{Strong electronic coupling}

We begin by considering an EET dimer with relatively strong electronic coupling $J_{12} = J_{21} \equiv J = 100\text{ cm}^{-1}$ and
energetic bias $E_1-E_2 = 100\text{ cm}^{-1}$, such that the splitting frequency, Eq.~(\ref{eq:wstar}),
is $\omega^* \approx 220\ {\rm cm}^{-1}$. The simulations are performed at room temperature,
$T = 300\text{ K}$, with a bath cutoff frequency of $\omega_c = 53\ {\rm cm}^{-1}$. 
The bath correlation time corresponding to this cutoff frequency is $\tau_c = \omega_c^{-1} = 100\ {\rm fs}$.
Since $\omega_c/J \approx 0.5$, this regime can be characterized as being weakly adiabatic.

The population of site 1 for these parameters, with the initial condition $\rho(0) = |1\rangle\langle 1| \exp(-\beta H_b) / Z$,
is shown in Fig.~\ref{fig:ishizaki4} for a reorganization energy varying over more than two order of magnitude.  
As alluded to in the introduction,
the Ehrenfest dynamics are seen to be qualitatively very good, though for intermediate reorganization energies, 
they show a significant deviation in the long-time population.
This behavior is to be contrasted with that observed using the LSC-IVR method\cite{tao10},
where the agreement regularly worsened with increasing reorganization energy (see Figs. 1 and 2 of Ref.~\onlinecite{tao10}).
The Ehrenfest and LSC-IVR approaches are almost identical for the simple Frenkel Hamiltonian considered here, 
the only difference being that the Ehrenfest approach treats the electronic degrees of freedom quantum mechanically
whereas LSC-IVR treats them by a Meyer-Miller mapping to classical oscillators\cite{mey79,sto97,tho99}, such that all dynamical degrees of
freedom are treated on equal footing. The results observed here indicate that such a consistent treatment does not
necessarily yield more accurate results, especially in regimes of strong reorganization energy. In light of these failures of
LSC-IVR and Ehrenfest approaches, it is crucial to notice that our RDM-Hybrid methodology yields nearly exact 
population dynamics for all values of
the reorganization energy, including an accurate treatment of long-time dynamics due to the quantum mechanical
treatment of high frequency environmental modes.

\begin{figure}[t]
\centering
\includegraphics[scale=0.4]{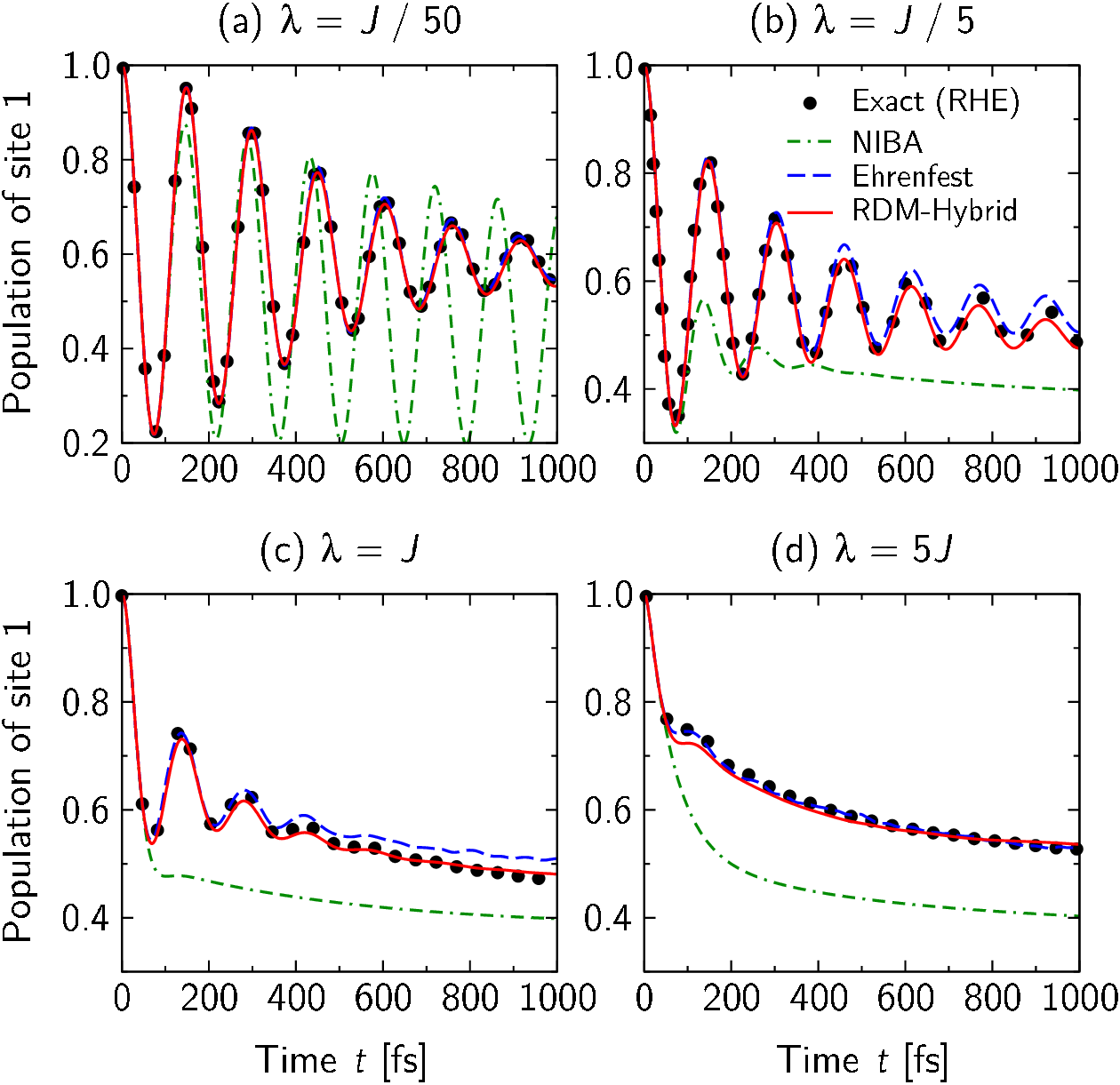}
\caption{The same as in Figure \ref{fig:ishizaki4}, but with $\omega_c = 11\ {\rm cm}^{-1}$
($\omega_c^{-1} = 500$~fs).
}
\label{fig:ishizaki6}
\end{figure}

Moving deeper into the adiabatic regime, we next consider the same EET dimer but with a smaller cutoff
frequency, $\omega_c = 11\ {\rm cm}^{-1}$ (longer bath correlation time, $\tau_c = 500$ fs), 
such that $\omega_c/J \approx 0.1$. Population dynamics are shown in 
Fig.~\ref{fig:ishizaki6} for the same range of reorganization energies as above.  
For this smaller cutoff frequency, Redfield dynamics have been shown to be inaccurate even for the 
smallest reorganization energy considered\cite{ish09_jcp1}. However, as discussed in Sec.~\ref{sec:intro},
quantum-classical methods are highly suitable in this strongly adiabatic regime. Indeed, the Ehrenfest
results presented here and the LSC-IVR results of Ref.~\onlinecite{tao10} are nearly exact. Again, the RDM-Hybrid
results are excellent, correcting the minor discrepancies seen in the long-time populations of the Ehrenfest 
dynamics. The RDM-Hybrid methodology naturally `tunes' itself to the
more accurate of its two composite methods. For example, in going from Fig.~\ref{fig:ishizaki4} to 
Fig.~\ref{fig:ishizaki6}, lowering the bath cutoff frequency further below the splitting frequency
results in treating a higher percentage of bath modes with Ehrenfest dynamics, the more accurate of the two methods in
this parameter regime. However, as we demonstrated in our earlier work,\cite{ber11_1} the RDM-Hybrid approach performs 
better than the sum of its parts and can also treat regimes where neither NIBA nor Ehrenfest 
dynamics alone would be suitable.

A recent work\cite{ish11} also presented Ehrenfest results similar to those shown in Figs.~\ref{fig:ishizaki4}
and \ref{fig:ishizaki6} but the back-reaction of the quantum system on the classical one, given at the end of 
Eq.~(\ref{eq:batheom2}), was neglected. Thus, the classical bath harmonic oscillators were isolated and provided only a 
fluctuating bias in the system equations of motion. This approximation is akin to well-known Haken-Strobl-Reinecker type
of master equation which yields equilibrium populations consistent with infinite temperature, i.e. both populations 
go to 1/2 in a dimer regardless of the bias, a deficient behavior observed in Ref.~\onlinecite{ish11}.
As can be seen in our Figs.~\ref{fig:ishizaki4} and \ref{fig:ishizaki6}, the full Ehrenfest treatment, including the 
back-reaction, always yields more accurate long-time populations (less than 1/2).

\subsubsection{Weak electronic coupling}

Although we argue that many of the interesting dynamical results in the recent EET literature can be ascribed 
to the system's adiabaticity, we
close our study of Ishizaki and Fleming's EET dimer by investigating a mildly non-adiabatic set of parameters, 
i.e. a dimer with weak electronic coupling, to demonstrate the flexibility of our RDM-Hybrid approach and continued
success of Ehrenfest dynamics. The parameters are the same as in the previous section, taking $\omega_c = 53\ {\rm cm}^{-1}$,
but now with $J = 20\ {\rm cm}^{-1}$, such that that adiabaticity ratio is $\omega_c/J \approx 3$.

\begin{figure}[t]
\centering
\includegraphics[scale=0.4]{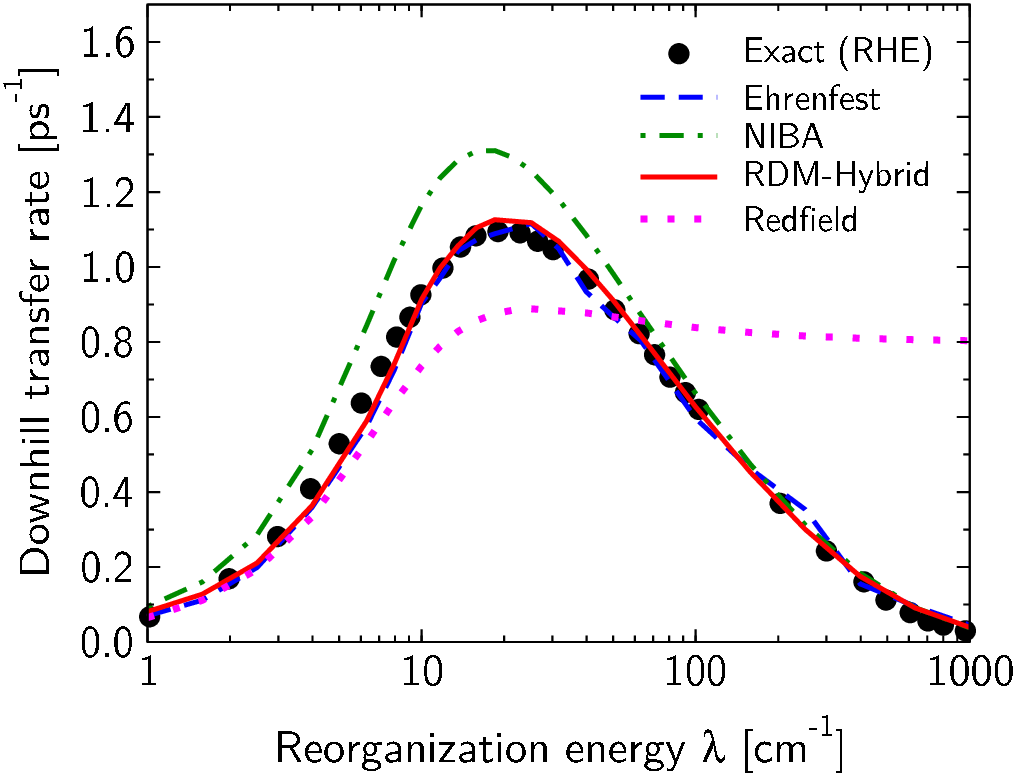}
\caption{Downhill energy transfer rates for an EET dimer in the weakly non-adiabatic regime ($\omega_c/J > 1$),
with $E_1-E_2 = 100\ {\rm cm}^{-1}$, $J = 20\ {\rm cm}^{-1}$, $\omega_c = 53\ {\rm cm}^{-1}$
($\omega_c^{-1} = 100$~fs), and $T = 300\ {\rm K}$.
}
\label{fig:ishizaki2}
\end{figure}

Rather than calculating population dynamics as above, we determine the downhill energy transfer rate as a function
of the reorganization energy, $\lambda$.  Due to the small electronic coupling, the population dynamics are
generally well described by an exponential decay and a simple fitting procedure yields the uphill and downhill rate
constants.  The results of the Ehrenfest and RDM-Hybrid approaches are shown in Fig.~\ref{fig:ishizaki2} and can be seen
to be in almost perfect agreement with the exact RHE results\cite{ish09_jcp2}. For completeness, we also present 
results obtained by conventional Redfield theory and the pure NIBA-like
equations derived in Sec.~\ref{sec:niba}. Redfield theory is perturbative in $\lambda kT/\omega_{c}^2$ and 
hence it would be expected to
break down when the reorganization energy $\lambda$ is on the order of $\omega_c^2/kT \approx 10\ {\rm cm}^{-1}$,
an order-of-magnitude prediction which is seen to hold unreasonably well
in Fig.~\ref{fig:ishizaki2}.  Furthermore, we point out that our NIBA-like
dynamical theory almost exactly reproduces the F{\" o}rster result presented in Ref.~\onlinecite{ish09_jcp2}, 
as one would expect due to their close formal relation, discussed at the end of Sec.~\ref{sec:niba}.

\subsection{Fenna-Matthews-Olson Complex}

\begin{figure}[t]
\centering
\includegraphics[scale=0.4]{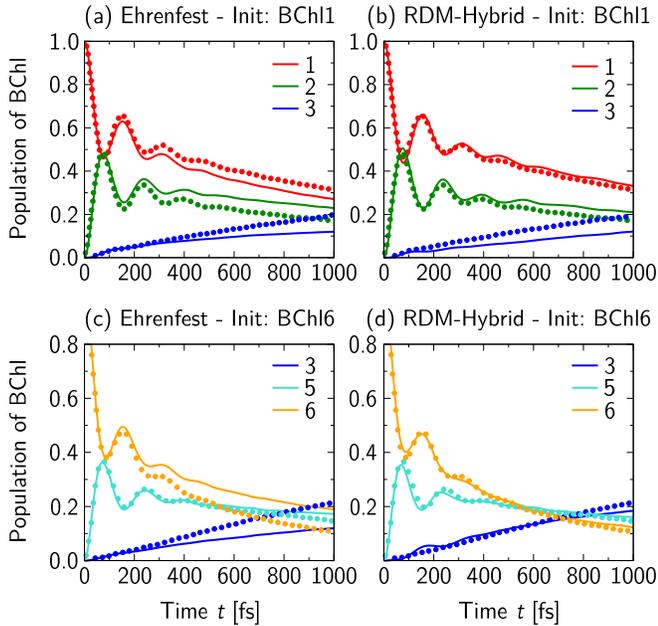}
\caption{Population dynamics of the FMO complex at $T=300$~K, with $\tau_c = \omega_c^{-1} = 166$~fs and bath initial
conditions sampled from the Wigner distribution (for both Ehrenfest and the present hybrid method).
The excitation
is initially localized to site 1 in panels (a) and (b) and to site 6 in panels (c) and (d).  
Ehrenfest [(a),(c)] and the RDM-Hybrid approach [(b),(d)] are compared to exact results obtained with the
reduced hierarchy equations (filled circles).
}
\label{fig:fmo4}
\end{figure}

We now proceed to the investigation of the Fenna-Matthews-Olsen (FMO) complex, which has been the subject of 
extensive experimental and theoretical investigation, especially pertaining to the
origin of long-lived quantum coherence. Although an eighth bacteriochlorophyll (BChl) chromophore has 
recently been identified\cite{sch11}, we consider the FMO model Hamiltonian of only seven BChl sites, for which 
numerically exact results exist\cite{ish09_pnas} and a variety of other approximate methods have been tested. The 
electronic Hamiltonian is taken from Ref.~\onlinecite{ado06} and all sites are assumed to have independent, identical 
baths characterized by a Debye spectral density, Eq.~\ref{eq:debye}, with $\lambda = 35\ {\rm cm}^{-1}$. The bath 
cutoff frequency and temperature will be varied throughout our investigation. Our initial results in this section 
will serve to demonstrate the accuracy of our RDM-Hybrid approach when treating such systems while the second 
part will use the method to study the effect of bath preparation on quantum coherent dynamics.

When moving from a dimer to a multi-site system, it becomes less obvious how to choose a characteristic system 
frequency for the switching frequency, $\omega^*$, required by our RDM-Hybrid method.  For the FMO system, we take
the switching frequency equal to the Rabi frequency of the initially excited site and its most strongly electronically-coupled neighbor.

\subsubsection{Comparison with existing results}\label{sssec:fmo}

We begin by comparing to the numerically exact results of Ishizaki and
Fleming\cite{ish09_pnas} and using the initial condition from that
work,
\begin{equation}
\rho(0) = |n_0\rangle\langle n_0| \exp(-\beta H_b) /Z\label{eq:unshifted},
\end{equation}
where the initial excitation site is taken to be $n_0 =$ 1 or 6 due to their proximity to the chlorophyll
baseplate. This sampling of the bath corresponds to a spectroscopic initial condition in which the
bath initial conditions are not equilibrated to the presence of the system.

\begin{figure}[t]
\centering
\includegraphics[scale=0.4]{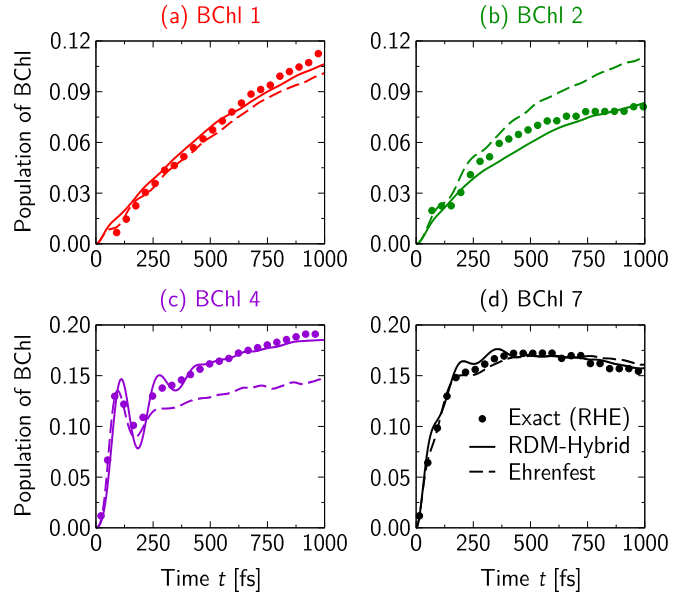}
\caption{Population dynamics of the remaining four BChl sites not depicted in Figs.~\ref{fig:fmo4}(c) and (d),
i.e. with site 6 initially excited.
}
\label{fig:fmo4_6_small}
\end{figure}

As discussed above, the switching frequency is set equal to the Rabi frequency of sites 1 and 2, 
yielding $\omega^* = 210\ {\rm cm}^{-1}$, or to the Rabi frequency of sites 5 and 6, 
yielding $\omega^* = 220\ {\rm cm}^{-1}$, for initial excitations at sites 1 or 6, respectively. 
In Fig.~\ref{fig:fmo4}, we consider the parameter set investigated by Ishizaki and Fleming for which Ehrenfest 
dynamics are expected to work best, namely a very slow, high temperature bath with $\tau_c = \omega_c^{-1} = 166$ fs 
and $T = 300$ K.  Despite the relatively weak reorganization energy, master equations that are perturbative in the 
system-bath coupling (such as Redfield theory) are unable to accurately reproduce these dynamics due to the 
system-bath adiabaticity.\footnote{T. C. Berkelbach and D. R. Reichman (unpublished)}  In contrast, the Ehrenfest 
approach performs very well for short times and accurately reproduces the coherence frequency, amplitude, and 
lifetime. However, just as in our above study of a two-level system, we see that the long-time populations deviate 
from the exact values, a flaw which is impressively remedied with our RDM-Hybrid approach, yielding excellent 
agreement overall. For example, by comparing Figs.~\ref{fig:fmo4}(c) and (d), we see that the RDM-Hybrid dynamics 
correctly reproduce the population inversion completely missed by the Ehrenfest dynamics.

For clarity, we only show population dynamics for three of the seven sites in Fig.~\ref{fig:fmo4}. However, 
the conclusions drawn are entirely unchanged for the four remaining sites with smaller populations,
as shown in Fig.~\ref{fig:fmo4_6_small} for the initial excitation $n_0 = 6$. 

\begin{figure}[t]
\centering
\includegraphics[scale=0.4]{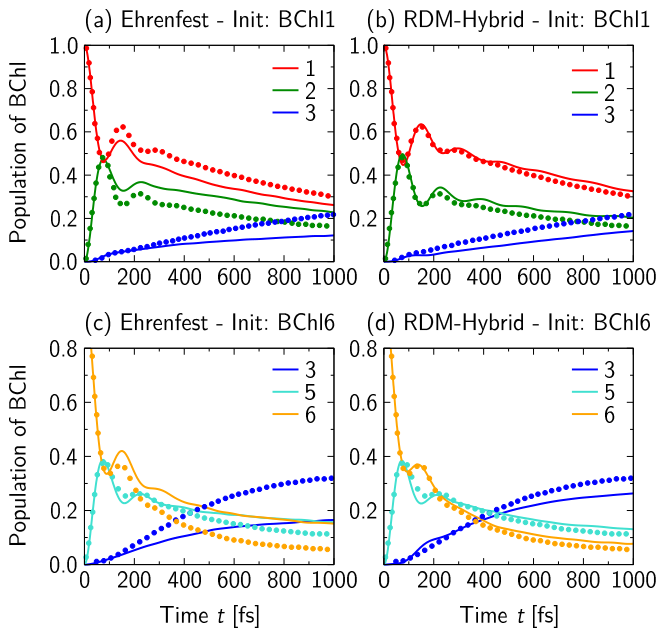}
\caption{The same as in Fig.~\ref{fig:fmo4}, but for the longer bath correlation time, $\tau_c = \omega_c^{-1} = 50$~fs
}
\label{fig:fmo3}
\end{figure}

By shortening the bath correlation time, we expect that the performance of Ehrenfest dynamics should degrade 
when compared to the exact result, since high frequency bath modes necessitate a quantum treatment.  Indeed, 
this expectation is realized in Fig.~\ref{fig:fmo3}, for which $\tau_c = \omega_c^{-1} = 50$ fs.  
Figures~\ref{fig:fmo3} (a) and (c) show that Ehrenfest dynamics again yield qualitatively accurate coherence 
lifetimes but incorrect
long-time populations, now worsened due to the short bath correlation time. In contrast to Ehrenfest dynamics, 
the quantum-mechanical treatment of high-frequency modes in the RDM-Hybrid methodology results in excellent 
performance, again exemplified by the population inversion in Fig.~\ref{fig:fmo3}(d).

\begin{figure}[t]
\centering
\includegraphics[scale=0.4]{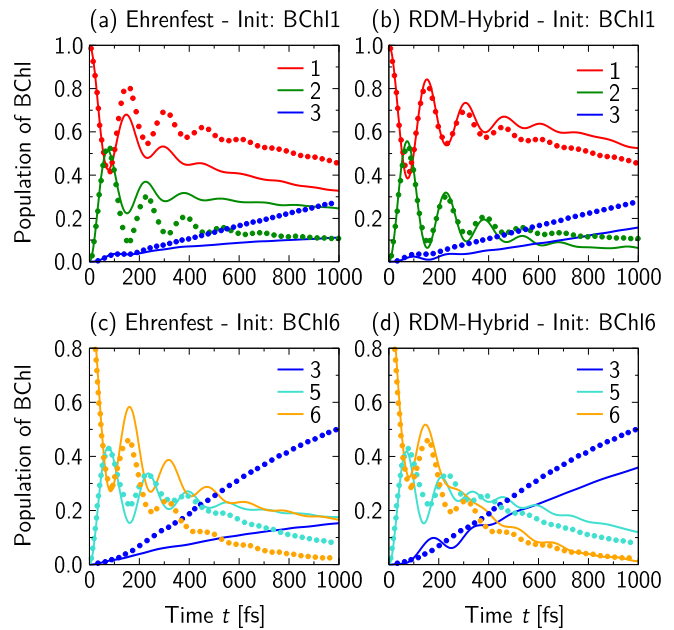}
\caption{Population dynamics of the FMO complex at $T=77$~K, with $\tau_c = \omega_c^{-1} = 50$~fs and bath initial
conditions sampled from the Wigner distribution (for both Ehrenfest and the present hybrid method).
The excitation
is initially localized to site 1 in panels (a) and (b) and to site 6 in panels (c) and (d).  
Ehrenfest [(a),(c)] and the RDM-Hybrid approach [(b),(d)] are compared to exact results obtained with the
reduced hierarchy equations (filled circles). 
}
\label{fig:fmo2}
\end{figure}

The final set of standard conditions considered here again has a short bath correlation
time $\tau_c = 50$ fs but at a reduced temperature, $T=77$ K corresponding to that at which some of the original experiments
observing quantum coherence were performed\cite{eng07}.  At this low temperature, both approximations
used in the present RDM-Hybrid implementation (NIBA and Ehrenfest) are known to worsen.  Nonetheless, the RDM-Hybrid
population dynamics shown in Fig.~\ref{fig:fmo2} are impressively good, and qualitatively much more accurate
than those of Ehrenfest dynamics.  However, both methodologies again make excellent prediction of the
coherence frequency and lifetime, which is the experimentally observed phenomenon that has generally
garnered the most attention.  It is also worth comparing again to the LSC-IVR work of Ref.~\onlinecite{tao10}, which
presented population dynamics for the FMO complex in the present parameter regime ($\tau_c = 50$ fs, $T = 77$ K).
The LSC-IVR calculation severely underestimate the coherence lifetime, yields some {\em negative} populations at long
times, and requires about $5 \times 10^4$ trajectories to achieve convergence -- almost two order of magnitude greater
than required by Ehrenfest of RDM-Hybrid approaches. 

\subsubsection{Effects of bath preparation on coherent transport}

The initial condition considered up to this point, Eq.~(\ref{eq:unshifted}),
is sometimes referred to as a `spectroscopic' preparation, as it corresponds to the physically
correct initial condition following a rapid excitation from the ground state in accordance with the Franck-Condon principle.  Therefore,
this initial condition is likely to most closely resemble the preparation realized in recent infrared spectroscopy experiments.
However in biological functioning, it is unlikely that this initial condition is physically correct for the dynamics of the
FMO complex, for example.  Recall that the excitation in the FMO complex is transferred in from the chlorophyll's baseplate.  This
long-range transfer is likely a non-adiabatic process and thus requires a fluctuation in the bath coordinates of the FMO acceptor
site leading to an excited state geometry, as in the traditional Marcus picture.  In light of this discussion it is clearly
worth investigating to what extent oscillatory population dynamics are modified by a non-spectroscopic initial condition and whether
the observed long-lasting quantum beating -- in both experiments and simulations -- is perhaps a product of unphysical
spectroscopic initial conditions.

Despite the recent interest surrounding quantum coherence in energy transfer molecules and materials, very similar
investigation of electronic and vibrational coherence in {\em electron} transfer reactions began almost twenty years ago
(see e.g. Refs.~\onlinecite{jea92,coa94,rei95,luc97}). Most related to our present investigation is the
theoretical work of Lucke et al.\cite{luc97}, who investigated the effects of initial bath preparation on the possible observance
of electronic coherence. The principal conclusion of their work was that absence of oscillatory population dynamics
does not necessarily imply dephasing-induced decoherence and that the way in which the bath is prepared can
suppress or enhance oscillations.

The RDM-Hybrid approach can very naturally treat arbitrary bath initial conditions and here
we consider two different choices for the initial bath density matrix, defined in Eqs.~(\ref{eq:bathic})-(\ref{eq:bathshift}).  What
we will term an `unshifted' initial condition has $\delta_n = 0$ for all sites $n$ (spectroscopic preparation),
whereas a `shifted' initial condition
has $\delta_{n_0} = -1$ where $n_0$ is the initially occupied system site and $\delta_n = 0$ for all other sites $n$.  Note that
within the RDM-Hybrid scheme, these shift parameters manifest in the sampling of the classical reservoir coordinates as well as
in the master equation memory kernel, Eq.~(\ref{eq:kernel}). 

\begin{figure}[t]
\centering
\includegraphics[scale=0.4]{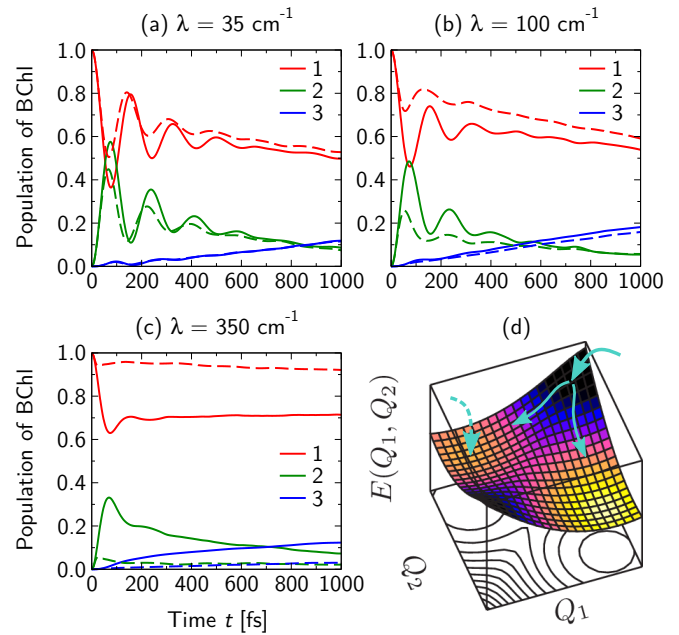}
\caption{Effects of initial bath preparation on the FMO population dynamics at $T=77$ K with
$\tau_c = \omega_c^{-1} = 166$ fs.  Solid lines depict the unshifted (spectroscopic) initial condition,
whereas dashed lines depict the shifted (solvated) initial condition, for a range of reorganization
energies, $\lambda$ (a)-(c).  Also shown is a schematic
diagram (d) of the lower energy adiabatic potential energy surface as a function of the generalized reaction coordinates,
$Q_1$ and $Q_2$, to which sites 1 and 2 are coupled.  Solid line arrows correspond to an unshifted
initial condition and subsequent excitation dynamics,
whereas the dotted line arrow corresponds to an initial condition shifted to the minimum of site 1, trapping the excitation.
}
\label{fig:fmo_shift}
\end{figure}

To best exemplify this phenomenon, we will consider a parameter regime for which such sensitivity to initial conditions
is expected to most strongly manifest,
namely for an adiabatic, low temperature bath. In particular, we consider $T=77$ K and $\tau_c = \omega_c^{-1} = 166$ fs 
with the reorganization energy to be varied.
In Fig.~\ref{fig:fmo_shift}(a), we show the effect of shifted initial conditions calculated with the RDM-Hybrid approach
for the standard reorganization energy $\lambda = 35\ {\rm cm}^{-1}$ and the excitation initially localized to 
the first BChl site.  Clearly,
the shifted initial condition yields population dynamics with a reduced oscillation amplitude, by about a factor of two.

Physically, the above effect occurs because the shifted initial condition brings the total system into a near-eigenstate,
in essence trapping the particle in a quasi-stationary state with no observable coherent oscillations.
On the contrary, with an unshifted initial condition, the total system is in a superposition of 
eigenstates, and coherent dynamics will be observed.

Further exploring this effect, we increase the reorganization energy in Figs.~\ref{fig:fmo_shift}(b) and (c) 
to $\lambda = 100\ {\rm cm}^{-1}$
and 350 cm$^{-1}$, respectively.  For $\lambda = 100\ {\rm cm}^{-1}$, the shifted initial condition can be seen to almost
completely suppress all population oscillations despite predicting almost identical population relaxation.  Increasing
further to the very strong $\lambda = 350\ {\rm cm}^{-1}$, we show that for the shifted initial condition, the bath completely traps
the excitation at the initial site, whereas for the unshifted initial condition, the bath can only trap the particle after
short-time coherence allows significant population transfer.
This situation is depicted schematically in Fig.~\ref{fig:fmo_shift}(d), where
the large, solid arrow marks the unshifted initial condition and the smaller solid arrows show the possible relaxation
into neighboring diabatic wells.  Relaxation into the diabatic well of site 2 corresponds to population transfer.
On the other hand, the large, dashed arrow marks an initial
condition shifted to the minimum of site 1, effectively trapping the excitation and preventing any population relaxation.
These two contrasting population relaxation behaviors can be seen clearly in the dynamics of Fig.~\ref{fig:fmo_shift}(c).

\section{Conclusions}\label{sec:conc}

To summarize, we have argued that the low frequency environmental motions present in many recent models of 
excitation energy transfer (EET) marks a significant
deviation from the validity regimes of popular existing methods, such as the Redfield and F\"{o}rster theories.  
In light of this observation, 
we have shown that quantum-classical approaches provide a simple and intuitive route towards more accurate 
modeling of EET in intermediate coupling regimes.  
Although the typical Ehrenfest method, which treats the electronic subsystem quantum mechanically
and the environment classically, yields reasonable results when compared to exact calculations 
for a wide range of EET Hamiltonians considered in the literature, its long-time populations suffer from 
well-known unrestricted energy flow, often yielding equilibrium populations corresponding
to an infinite temperature result, i.e. equal population of all sites.

To alleviate this problem, we have employed our recently developed RDM-Hybrid algorithm\cite{ber11_1}, 
which extends the usual Ehrenfest method by including high-frequency environmental modes into a quantum ``core.''  
Importantly, for pure system properties, one need only calculate the reduced density matrix averaged over the 
quantum core, which can be done approximately but accurately for the high-frequency modes included.
In turn, the remaining ``reservoir'' modes are treated classically and the usual mean-field coupling exists 
between the system-core and reservoir degrees of freedom.  Such an approach yields excellent results and 
can be applied without modification to a great variety of system-bath Hamiltonians, extending well beyond the 
domain of EET parameters.

In addition to the very favorable comparison with existing results, the RDM-Hybrid method was also employed for a novel
investigation of initial bath preparation and its effects on subsequent population dynamics. In particular, 
we showed that the experimentally relevant spectroscopic initial conditions
often employed in calculations may be partly responsible for the unexpectedly long-lived quantum 
coherence. The degree to which this behavior manifests in biological functioning
will be sensitive to the way in which excitations enter FMO and related complexes, a topic which has 
received almost no attention in the literature but is surely deserving of further investigation.

Lastly, we point out that whereas most existing exact methodologies,
including influence functional-based path-integral methods and the reduced hierarchy equations, are reliant
on harmonic bath degrees of freedom, the Ehrenfest approximation
is applicable for systems with generically anharmonic degrees of freedom.
Thus, the work performed here marks an important step towards the treatment of EET in
realistic molecular systems and could easily incorporate low-frequency anharmonic modes.  High-frequency
anharmonicities could also be treated, but would require numerical evaluation of the memory kernel using e.g.
semi-classical methods\cite{ner93,shi04}.  Such work is currently in progress. 

\begin{acknowledgments}
The authors thank Aaron Kelly for useful insights and suggestions on this manuscript. T.C.B.\ was 
supported by the Department of Energy Office of Science Graduate Fellowship Program (DOE SCGF),
administered by ORISE-ORAU under Contract No.~DE-AC05-06OR23100 and D.R.R.\ was supported by the National Science
Foundation under Grant No.~CHE-0719089.
\end{acknowledgments}

%\bibliography{relaxation}

\begin{thebibliography}{45}%
\makeatletter
\providecommand \@ifxundefined [1]{%
 \@ifx{#1\undefined}
}%
\providecommand \@ifnum [1]{%
 \ifnum #1\expandafter \@firstoftwo
 \else \expandafter \@secondoftwo
 \fi
}%
\providecommand \@ifx [1]{%
 \ifx #1\expandafter \@firstoftwo
 \else \expandafter \@secondoftwo
 \fi
}%
\providecommand \natexlab [1]{#1}%
\providecommand \enquote  [1]{``#1''}%
\providecommand \bibnamefont  [1]{#1}%
\providecommand \bibfnamefont [1]{#1}%
\providecommand \citenamefont [1]{#1}%
\providecommand \href@noop [0]{\@secondoftwo}%
\providecommand \href [0]{\begingroup \@sanitize@url \@href}%
\providecommand \@href[1]{\@@startlink{#1}\@@href}%
\providecommand \@@href[1]{\endgroup#1\@@endlink}%
\providecommand \@sanitize@url [0]{\catcode `\\12\catcode `\$12\catcode
  `\&12\catcode `\#12\catcode `\^12\catcode `\_12\catcode `\%12\relax}%
\providecommand \@@startlink[1]{}%
\providecommand \@@endlink[0]{}%
\providecommand \url  [0]{\begingroup\@sanitize@url \@url }%
\providecommand \@url [1]{\endgroup\@href {#1}{\urlprefix }}%
\providecommand \urlprefix  [0]{URL }%
\providecommand \Eprint [0]{\href }%
\providecommand \doibase [0]{http://dx.doi.org/}%
\providecommand \selectlanguage [0]{\@gobble}%
\providecommand \bibinfo  [0]{\@secondoftwo}%
\providecommand \bibfield  [0]{\@secondoftwo}%
\providecommand \translation [1]{[#1]}%
\providecommand \BibitemOpen [0]{}%
\providecommand \bibitemStop [0]{}%
\providecommand \bibitemNoStop [0]{.\EOS\space}%
\providecommand \EOS [0]{\spacefactor3000\relax}%
\providecommand \BibitemShut  [1]{\csname bibitem#1\endcsname}%
\let\auto@bib@innerbib\@empty
%</preamble>
\bibitem [{\citenamefont {Engel}\ \emph {et~al.}(2007)\citenamefont {Engel},
  \citenamefont {Calhoun}, \citenamefont {Read}, \citenamefont {Ahn},
  \citenamefont {Mancal}, \citenamefont {Cheng}, \citenamefont {Blankenship},\
  and\ \citenamefont {Fleming}}]{eng07}%
  \BibitemOpen
  \bibfield  {author} {\bibinfo {author} {\bibfnamefont {G.~S.}\ \bibnamefont
  {Engel}}, \bibinfo {author} {\bibfnamefont {T.~R.}\ \bibnamefont {Calhoun}},
  \bibinfo {author} {\bibfnamefont {E.~L.}\ \bibnamefont {Read}}, \bibinfo
  {author} {\bibfnamefont {T.-K.}\ \bibnamefont {Ahn}}, \bibinfo {author}
  {\bibfnamefont {T.}~\bibnamefont {Mancal}}, \bibinfo {author} {\bibfnamefont
  {Y.-C.}\ \bibnamefont {Cheng}}, \bibinfo {author} {\bibfnamefont {R.~E.}\
  \bibnamefont {Blankenship}}, \ and\ \bibinfo {author} {\bibfnamefont {G.~R.}\
  \bibnamefont {Fleming}},\ }\href@noop {} {\bibfield  {journal} {\bibinfo
  {journal} {Nature}\ }\textbf {\bibinfo {volume} {446}},\ \bibinfo {pages}
  {782} (\bibinfo {year} {2007})}\BibitemShut {NoStop}%
\bibitem [{\citenamefont {Lee}, \citenamefont {Cheng},\ and\ \citenamefont
  {Fleming}(2007)}]{lee07}%
  \BibitemOpen
  \bibfield  {author} {\bibinfo {author} {\bibfnamefont {H.}~\bibnamefont
  {Lee}}, \bibinfo {author} {\bibfnamefont {Y.-C.}\ \bibnamefont {Cheng}}, \
  and\ \bibinfo {author} {\bibfnamefont {G.~R.}\ \bibnamefont {Fleming}},\
  }\href@noop {} {\bibfield  {journal} {\bibinfo  {journal} {Science}\ }\textbf
  {\bibinfo {volume} {316}},\ \bibinfo {pages} {1462} (\bibinfo {year}
  {2007})}\BibitemShut {NoStop}%
\bibitem [{\citenamefont {Collini}\ and\ \citenamefont
  {Scholes}(2009{\natexlab{a}})}]{col09_1}%
  \BibitemOpen
  \bibfield  {author} {\bibinfo {author} {\bibfnamefont {E.}~\bibnamefont
  {Collini}}\ and\ \bibinfo {author} {\bibfnamefont {G.~D.}\ \bibnamefont
  {Scholes}},\ }\href@noop {} {\bibfield  {journal} {\bibinfo  {journal}
  {Science}\ }\textbf {\bibinfo {volume} {323}},\ \bibinfo {pages} {369}
  (\bibinfo {year} {2009}{\natexlab{a}})}\BibitemShut {NoStop}%
\bibitem [{\citenamefont {Collini}\ and\ \citenamefont
  {Scholes}(2009{\natexlab{b}})}]{col09_2}%
  \BibitemOpen
  \bibfield  {author} {\bibinfo {author} {\bibfnamefont {E.}~\bibnamefont
  {Collini}}\ and\ \bibinfo {author} {\bibfnamefont {G.~D.}\ \bibnamefont
  {Scholes}},\ }\href@noop {} {\bibfield  {journal} {\bibinfo  {journal} {J.
  Phys. Chem. A}\ }\textbf {\bibinfo {volume} {113}},\ \bibinfo {pages} {4223}
  (\bibinfo {year} {2009}{\natexlab{b}})}\BibitemShut {NoStop}%
\bibitem [{\citenamefont {Cheng}\ and\ \citenamefont {Fleming}(2009)}]{che09}%
  \BibitemOpen
  \bibfield  {author} {\bibinfo {author} {\bibfnamefont {Y.-C.}\ \bibnamefont
  {Cheng}}\ and\ \bibinfo {author} {\bibfnamefont {G.~R.}\ \bibnamefont
  {Fleming}},\ }\href@noop {} {\bibfield  {journal} {\bibinfo  {journal} {Ann.
  Rev. Phys. Chem.}\ }\textbf {\bibinfo {volume} {60}},\ \bibinfo {pages} {241}
  (\bibinfo {year} {2009})}\BibitemShut {NoStop}%
\bibitem [{\citenamefont {Ishizaki}\ \emph {et~al.}(2010)\citenamefont
  {Ishizaki}, \citenamefont {Calhoun}, \citenamefont {Schlau-Cohen},\ and\
  \citenamefont {Fleming}}]{ish10}%
  \BibitemOpen
  \bibfield  {author} {\bibinfo {author} {\bibfnamefont {A.}~\bibnamefont
  {Ishizaki}}, \bibinfo {author} {\bibfnamefont {T.~R.}\ \bibnamefont
  {Calhoun}}, \bibinfo {author} {\bibfnamefont {G.~S.}\ \bibnamefont
  {Schlau-Cohen}}, \ and\ \bibinfo {author} {\bibfnamefont {G.~R.}\
  \bibnamefont {Fleming}},\ }\href@noop {} {\bibfield  {journal} {\bibinfo
  {journal} {Phys. Chem. Chem. Phys.}\ }\textbf {\bibinfo {volume} {12}},\
  \bibinfo {pages} {7319} (\bibinfo {year} {2010})}\BibitemShut {NoStop}%
\bibitem [{\citenamefont {Yang}\ and\ \citenamefont {Fleming}(2002)}]{yan02}%
  \BibitemOpen
  \bibfield  {author} {\bibinfo {author} {\bibfnamefont {M.}~\bibnamefont
  {Yang}}\ and\ \bibinfo {author} {\bibfnamefont {G.~R.}\ \bibnamefont
  {Fleming}},\ }\href@noop {} {\bibfield  {journal} {\bibinfo  {journal} {Chem.
  Phys.}\ }\textbf {\bibinfo {volume} {282}},\ \bibinfo {pages} {163} (\bibinfo
  {year} {2002})}\BibitemShut {NoStop}%
\bibitem [{\citenamefont {Mohseni}\ \emph {et~al.}(2008)\citenamefont
  {Mohseni}, \citenamefont {Rebentrost}, \citenamefont {Lloyd},\ and\
  \citenamefont {Aspuru-Guzik}}]{moh08}%
  \BibitemOpen
  \bibfield  {author} {\bibinfo {author} {\bibfnamefont {M.}~\bibnamefont
  {Mohseni}}, \bibinfo {author} {\bibfnamefont {P.}~\bibnamefont {Rebentrost}},
  \bibinfo {author} {\bibfnamefont {S.}~\bibnamefont {Lloyd}}, \ and\ \bibinfo
  {author} {\bibfnamefont {A.}~\bibnamefont {Aspuru-Guzik}},\ }\href@noop {}
  {\bibfield  {journal} {\bibinfo  {journal} {J. Chem. Phys.}\ }\textbf
  {\bibinfo {volume} {129}},\ \bibinfo {pages} {174106} (\bibinfo {year}
  {2008})}\BibitemShut {NoStop}%
\bibitem [{\citenamefont {Rebentrost}, \citenamefont {Chakraborty},\ and\
  \citenamefont {Aspuru-Guzik}(2009)}]{reb09}%
  \BibitemOpen
  \bibfield  {author} {\bibinfo {author} {\bibfnamefont {P.}~\bibnamefont
  {Rebentrost}}, \bibinfo {author} {\bibfnamefont {R.}~\bibnamefont
  {Chakraborty}}, \ and\ \bibinfo {author} {\bibfnamefont {A.}~\bibnamefont
  {Aspuru-Guzik}},\ }\href@noop {} {\bibfield  {journal} {\bibinfo  {journal}
  {J. Chem. Phys.}\ }\textbf {\bibinfo {volume} {131}},\ \bibinfo {pages}
  {184102} (\bibinfo {year} {2009})}\BibitemShut {NoStop}%
\bibitem [{\citenamefont {Ishizaki}\ and\ \citenamefont
  {Fleming}(2009{\natexlab{a}})}]{ish09_jcp1}%
  \BibitemOpen
  \bibfield  {author} {\bibinfo {author} {\bibfnamefont {A.}~\bibnamefont
  {Ishizaki}}\ and\ \bibinfo {author} {\bibfnamefont {G.~R.}\ \bibnamefont
  {Fleming}},\ }\href@noop {} {\bibfield  {journal} {\bibinfo  {journal} {J.
  Chem. Phys.}\ }\textbf {\bibinfo {volume} {130}},\ \bibinfo {pages} {234110}
  (\bibinfo {year} {2009}{\natexlab{a}})}\BibitemShut {NoStop}%
\bibitem [{\citenamefont {Ishizaki}\ and\ \citenamefont
  {Fleming}(2009{\natexlab{b}})}]{ish09_jcp2}%
  \BibitemOpen
  \bibfield  {author} {\bibinfo {author} {\bibfnamefont {A.}~\bibnamefont
  {Ishizaki}}\ and\ \bibinfo {author} {\bibfnamefont {G.~R.}\ \bibnamefont
  {Fleming}},\ }\href@noop {} {\bibfield  {journal} {\bibinfo  {journal} {J.
  Chem. Phys.}\ }\textbf {\bibinfo {volume} {130}},\ \bibinfo {pages} {234111}
  (\bibinfo {year} {2009}{\natexlab{b}})}\BibitemShut {NoStop}%
\bibitem [{\citenamefont {Ishizaki}\ and\ \citenamefont
  {Fleming}(2009{\natexlab{c}})}]{ish09_pnas}%
  \BibitemOpen
  \bibfield  {author} {\bibinfo {author} {\bibfnamefont {A.}~\bibnamefont
  {Ishizaki}}\ and\ \bibinfo {author} {\bibfnamefont {G.~R.}\ \bibnamefont
  {Fleming}},\ }\href@noop {} {\bibfield  {journal} {\bibinfo  {journal} {Proc.
  Natl. Acad. Sci.}\ }\textbf {\bibinfo {volume} {106}},\ \bibinfo {pages}
  {17255} (\bibinfo {year} {2009}{\natexlab{c}})}\BibitemShut {NoStop}%
\bibitem [{\citenamefont {Jang}\ \emph {et~al.}(2008)\citenamefont {Jang},
  \citenamefont {Cheng}, \citenamefont {Reichman},\ and\ \citenamefont
  {Eaves}}]{jan08}%
  \BibitemOpen
  \bibfield  {author} {\bibinfo {author} {\bibfnamefont {S.}~\bibnamefont
  {Jang}}, \bibinfo {author} {\bibfnamefont {Y.-C.}\ \bibnamefont {Cheng}},
  \bibinfo {author} {\bibfnamefont {D.~R.}\ \bibnamefont {Reichman}}, \ and\
  \bibinfo {author} {\bibfnamefont {J.~D.}\ \bibnamefont {Eaves}},\ }\href@noop
  {} {\bibfield  {journal} {\bibinfo  {journal} {J. Chem. Phys.}\ }\textbf
  {\bibinfo {volume} {129}},\ \bibinfo {pages} {101104} (\bibinfo {year}
  {2008})}\BibitemShut {NoStop}%
\bibitem [{\citenamefont {Huo}\ and\ \citenamefont {Coker}(2010)}]{huo10}%
  \BibitemOpen
  \bibfield  {author} {\bibinfo {author} {\bibfnamefont {P.}~\bibnamefont
  {Huo}}\ and\ \bibinfo {author} {\bibfnamefont {D.~F.}\ \bibnamefont
  {Coker}},\ }\href@noop {} {\bibfield  {journal} {\bibinfo  {journal} {J.
  Chem. Phys.}\ }\textbf {\bibinfo {volume} {133}},\ \bibinfo {pages} {184108}
  (\bibinfo {year} {2010})}\BibitemShut {NoStop}%
\bibitem [{\citenamefont {Tao}\ and\ \citenamefont {Miller}(2010)}]{tao10}%
  \BibitemOpen
  \bibfield  {author} {\bibinfo {author} {\bibfnamefont {G.}~\bibnamefont
  {Tao}}\ and\ \bibinfo {author} {\bibfnamefont {W.~H.}\ \bibnamefont
  {Miller}},\ }\href@noop {} {\bibfield  {journal} {\bibinfo  {journal} {J.
  Phys. Chem. Lett.}\ }\textbf {\bibinfo {volume} {1}},\ \bibinfo {pages} {891}
  (\bibinfo {year} {2010})}\BibitemShut {NoStop}%
\bibitem [{\citenamefont {Jang}(2011)}]{jan11}%
  \BibitemOpen
  \bibfield  {author} {\bibinfo {author} {\bibfnamefont {S.}~\bibnamefont
  {Jang}},\ }\href@noop {} {\bibfield  {journal} {\bibinfo  {journal} {J. Chem.
  Phys.}\ }\textbf {\bibinfo {volume} {135}},\ \bibinfo {pages} {034105}
  (\bibinfo {year} {2011})}\BibitemShut {NoStop}%
\bibitem [{\citenamefont {Kelly}\ and\ \citenamefont {Rhee}(2011)}]{kel11}%
  \BibitemOpen
  \bibfield  {author} {\bibinfo {author} {\bibfnamefont {A.}~\bibnamefont
  {Kelly}}\ and\ \bibinfo {author} {\bibfnamefont {Y.~M.}\ \bibnamefont
  {Rhee}},\ }\href@noop {} {\bibfield  {journal} {\bibinfo  {journal} {J. Phys.
  Chem. Lett.}\ }\textbf {\bibinfo {volume} {2}},\ \bibinfo {pages} {808}
  (\bibinfo {year} {2011})}\BibitemShut {NoStop}%
\bibitem [{\citenamefont {Renaud}, \citenamefont {Ratner},\ and\ \citenamefont
  {Mujica}(2011)}]{ren11}%
  \BibitemOpen
  \bibfield  {author} {\bibinfo {author} {\bibfnamefont {N.}~\bibnamefont
  {Renaud}}, \bibinfo {author} {\bibfnamefont {M.~A.}\ \bibnamefont {Ratner}},
  \ and\ \bibinfo {author} {\bibfnamefont {V.~A.}\ \bibnamefont {Mujica}},\
  }\href@noop {} {\bibfield  {journal} {\bibinfo  {journal} {J. Chem. Phys.}\
  }\textbf {\bibinfo {volume} {135}},\ \bibinfo {pages} {075102} (\bibinfo
  {year} {2011})}\BibitemShut {NoStop}%
\bibitem [{\citenamefont {Nalbach}\ \emph {et~al.}(2011)\citenamefont
  {Nalbach}, \citenamefont {Ishizaki}, \citenamefont {Fleming},\ and\
  \citenamefont {Thorwart}}]{nal11}%
  \BibitemOpen
  \bibfield  {author} {\bibinfo {author} {\bibfnamefont {P.}~\bibnamefont
  {Nalbach}}, \bibinfo {author} {\bibfnamefont {A.}~\bibnamefont {Ishizaki}},
  \bibinfo {author} {\bibfnamefont {G.~R.}\ \bibnamefont {Fleming}}, \ and\
  \bibinfo {author} {\bibfnamefont {M.}~\bibnamefont {Thorwart}},\ }\href@noop
  {} {\bibfield  {journal} {\bibinfo  {journal} {New J. Phys.}\ }\textbf
  {\bibinfo {volume} {13}},\ \bibinfo {pages} {063040} (\bibinfo {year}
  {2011})}\BibitemShut {NoStop}%
\bibitem [{\citenamefont {Olbrich}\ \emph
  {et~al.}(2011{\natexlab{a}})\citenamefont {Olbrich}, \citenamefont {Stru{\"
  u}mpfer}, \citenamefont {Schulten},\ and\ \citenamefont
  {Kleinekathoefer}}]{olb11_jpcl}%
  \BibitemOpen
  \bibfield  {author} {\bibinfo {author} {\bibfnamefont {C.}~\bibnamefont
  {Olbrich}}, \bibinfo {author} {\bibfnamefont {J.}~\bibnamefont {Stru{\"
  u}mpfer}}, \bibinfo {author} {\bibfnamefont {K.}~\bibnamefont {Schulten}}, \
  and\ \bibinfo {author} {\bibfnamefont {U.}~\bibnamefont {Kleinekathoefer}},\
  }\href@noop {} {\bibfield  {journal} {\bibinfo  {journal} {J. Phys. Chem.
  Lett.}\ }\textbf {\bibinfo {volume} {2}},\ \bibinfo {pages} {1771} (\bibinfo
  {year} {2011}{\natexlab{a}})}\BibitemShut {NoStop}%
\bibitem [{\citenamefont {Olbrich}\ \emph
  {et~al.}(2011{\natexlab{b}})\citenamefont {Olbrich}, \citenamefont {Jansen},
  \citenamefont {Liebers}, \citenamefont {Aghtar}, \citenamefont {Stru{\"
  u}mpfer}, \citenamefont {Schulten}, \citenamefont {Knoester},\ and\
  \citenamefont {Kleinekathoefer}}]{olb11_jpcb}%
  \BibitemOpen
  \bibfield  {author} {\bibinfo {author} {\bibfnamefont {C.}~\bibnamefont
  {Olbrich}}, \bibinfo {author} {\bibfnamefont {T.~L.~C.}\ \bibnamefont
  {Jansen}}, \bibinfo {author} {\bibfnamefont {J.}~\bibnamefont {Liebers}},
  \bibinfo {author} {\bibfnamefont {M.}~\bibnamefont {Aghtar}}, \bibinfo
  {author} {\bibfnamefont {J.}~\bibnamefont {Stru{\" u}mpfer}}, \bibinfo
  {author} {\bibfnamefont {K.}~\bibnamefont {Schulten}}, \bibinfo {author}
  {\bibfnamefont {J.}~\bibnamefont {Knoester}}, \ and\ \bibinfo {author}
  {\bibfnamefont {U.}~\bibnamefont {Kleinekathoefer}},\ }\href@noop {}
  {\bibfield  {journal} {\bibinfo  {journal} {J. Phys. Chem. B}\ }\textbf
  {\bibinfo {volume} {115}},\ \bibinfo {pages} {8609} (\bibinfo {year}
  {2011}{\natexlab{b}})}\BibitemShut {NoStop}%
\bibitem [{\citenamefont {Redfield}(1965)}]{red65}%
  \BibitemOpen
  \bibfield  {author} {\bibinfo {author} {\bibfnamefont {A.~G.}\ \bibnamefont
  {Redfield}},\ }\href@noop {} {\bibfield  {journal} {\bibinfo  {journal} {Adv.
  Magn. Reson.}\ }\textbf {\bibinfo {volume} {1}},\ \bibinfo {pages} {1}
  (\bibinfo {year} {1965})}\BibitemShut {NoStop}%
\bibitem [{\citenamefont {F{\" o}rster}(1953)}]{for53}%
  \BibitemOpen
  \bibfield  {author} {\bibinfo {author} {\bibfnamefont {T.}~\bibnamefont {F{\"
  o}rster}},\ }\href@noop {} {\bibfield  {journal} {\bibinfo  {journal}
  {Discuss. Faraday Soc.}\ }\textbf {\bibinfo {volume} {27}},\ \bibinfo {pages}
  {7} (\bibinfo {year} {1953})}\BibitemShut {NoStop}%
\bibitem [{\citenamefont {Ishizaki}\ and\ \citenamefont
  {Fleming}(2011)}]{ish11}%
  \BibitemOpen
  \bibfield  {author} {\bibinfo {author} {\bibfnamefont {A.}~\bibnamefont
  {Ishizaki}}\ and\ \bibinfo {author} {\bibfnamefont {G.~R.}\ \bibnamefont
  {Fleming}},\ }\href@noop {} {\bibfield  {journal} {\bibinfo  {journal} {J.
  Phys. Chem. B}\ }\textbf {\bibinfo {volume} {115}},\ \bibinfo {pages} {6227}
  (\bibinfo {year} {2011})}\BibitemShut {NoStop}%
\bibitem [{\citenamefont {Berkelbach}, \citenamefont {Reichman},\ and\
  \citenamefont {Markland}(2011)}]{ber11_1}%
  \BibitemOpen
  \bibfield  {author} {\bibinfo {author} {\bibfnamefont {T.~C.}\ \bibnamefont
  {Berkelbach}}, \bibinfo {author} {\bibfnamefont {D.~R.}\ \bibnamefont
  {Reichman}}, \ and\ \bibinfo {author} {\bibfnamefont {T.~E.}\ \bibnamefont
  {Markland}},\ }\href@noop {} {} (\bibinfo {year} {2011}),\ \Eprint
  {http://arxiv.org/abs/arXiv:1110.0490} {arXiv:1110.0490} \BibitemShut
  {NoStop}%
\bibitem [{\citenamefont {Weiss}(2008)}]{wei08}%
  \BibitemOpen
  \bibfield  {author} {\bibinfo {author} {\bibfnamefont {U.}~\bibnamefont
  {Weiss}},\ }\href@noop {} {\emph {\bibinfo {title} {Quantum Dissipative
  Systems}}}\ (\bibinfo  {publisher} {World Scientific Publishing},\ \bibinfo
  {year} {2008})\BibitemShut {NoStop}%
\bibitem [{\citenamefont {Leggett}\ \emph {et~al.}(1987)\citenamefont
  {Leggett}, \citenamefont {Chakravarty}, \citenamefont {Dorsey}, \citenamefont
  {Fisher}, \citenamefont {Garg},\ and\ \citenamefont {Zwerger}}]{leg87}%
  \BibitemOpen
  \bibfield  {author} {\bibinfo {author} {\bibfnamefont {A.~J.}\ \bibnamefont
  {Leggett}}, \bibinfo {author} {\bibfnamefont {S.}~\bibnamefont
  {Chakravarty}}, \bibinfo {author} {\bibfnamefont {A.~T.}\ \bibnamefont
  {Dorsey}}, \bibinfo {author} {\bibfnamefont {M.~P.~A.}\ \bibnamefont
  {Fisher}}, \bibinfo {author} {\bibfnamefont {A.}~\bibnamefont {Garg}}, \ and\
  \bibinfo {author} {\bibfnamefont {W.}~\bibnamefont {Zwerger}},\ }\href@noop
  {} {\bibfield  {journal} {\bibinfo  {journal} {Rev. Mod. Phys.}\ }\textbf
  {\bibinfo {volume} {59}},\ \bibinfo {pages} {1} (\bibinfo {year}
  {1987})}\BibitemShut {NoStop}%
\bibitem [{\citenamefont {Egger}, \citenamefont {Mak},\ and\ \citenamefont
  {Weiss}(1994)}]{egg94}%
  \BibitemOpen
  \bibfield  {author} {\bibinfo {author} {\bibfnamefont {R.}~\bibnamefont
  {Egger}}, \bibinfo {author} {\bibfnamefont {C.~H.}\ \bibnamefont {Mak}}, \
  and\ \bibinfo {author} {\bibfnamefont {U.}~\bibnamefont {Weiss}},\
  }\href@noop {} {\bibfield  {journal} {\bibinfo  {journal} {Phys. Rev. E}\
  }\textbf {\bibinfo {volume} {50}},\ \bibinfo {pages} {R655} (\bibinfo {year}
  {1994})}\BibitemShut {NoStop}%
\bibitem [{\citenamefont {Hu}\ and\ \citenamefont {Mukamel}(1989)}]{hu89}%
  \BibitemOpen
  \bibfield  {author} {\bibinfo {author} {\bibfnamefont {Y.~M.}\ \bibnamefont
  {Hu}}\ and\ \bibinfo {author} {\bibfnamefont {S.}~\bibnamefont {Mukamel}},\
  }\href@noop {} {\bibfield  {journal} {\bibinfo  {journal} {J. Chem. Phys.}\
  }\textbf {\bibinfo {volume} {91}},\ \bibinfo {pages} {6973} (\bibinfo {year}
  {1989})}\BibitemShut {NoStop}%
\bibitem [{\citenamefont {Golosov}\ and\ \citenamefont
  {Reichman}(2001)}]{gol01}%
  \BibitemOpen
  \bibfield  {author} {\bibinfo {author} {\bibfnamefont {A.~A.}\ \bibnamefont
  {Golosov}}\ and\ \bibinfo {author} {\bibfnamefont {D.~R.}\ \bibnamefont
  {Reichman}},\ }\href@noop {} {\bibfield  {journal} {\bibinfo  {journal} {J.
  Chem. Phys.}\ }\textbf {\bibinfo {volume} {115}},\ \bibinfo {pages} {9848}
  (\bibinfo {year} {2001})}\BibitemShut {NoStop}%
\bibitem [{\citenamefont {Breuer}\ and\ \citenamefont
  {Petruccione}(2002)}]{bre02}%
  \BibitemOpen
  \bibfield  {author} {\bibinfo {author} {\bibfnamefont {H.-P.}\ \bibnamefont
  {Breuer}}\ and\ \bibinfo {author} {\bibfnamefont {F.}~\bibnamefont
  {Petruccione}},\ }\href@noop {} {\emph {\bibinfo {title} {The Theory of Open
  Quantum Systems}}}\ (\bibinfo  {publisher} {Oxford University Press},\
  \bibinfo {year} {2002})\BibitemShut {NoStop}%
\bibitem [{\citenamefont {Takagahara}, \citenamefont {Hanamura},\ and\
  \citenamefont {Kubo}(1977)}]{tak77}%
  \BibitemOpen
  \bibfield  {author} {\bibinfo {author} {\bibfnamefont {T.}~\bibnamefont
  {Takagahara}}, \bibinfo {author} {\bibfnamefont {E.}~\bibnamefont
  {Hanamura}}, \ and\ \bibinfo {author} {\bibfnamefont {R.}~\bibnamefont
  {Kubo}},\ }\href@noop {} {\bibfield  {journal} {\bibinfo  {journal} {J. Phys.
  Soc. Jpn.}\ }\textbf {\bibinfo {volume} {43}},\ \bibinfo {pages} {8111}
  (\bibinfo {year} {1977})}\BibitemShut {NoStop}%
\bibitem [{\citenamefont {Tanimura}\ and\ \citenamefont {Kubo}(1989)}]{tan89}%
  \BibitemOpen
  \bibfield  {author} {\bibinfo {author} {\bibfnamefont {Y.}~\bibnamefont
  {Tanimura}}\ and\ \bibinfo {author} {\bibfnamefont {R.}~\bibnamefont
  {Kubo}},\ }\href@noop {} {\bibfield  {journal} {\bibinfo  {journal} {J. Phys.
  Soc. Jpn.}\ }\textbf {\bibinfo {volume} {58}},\ \bibinfo {pages} {101}
  (\bibinfo {year} {1989})}\BibitemShut {NoStop}%
\bibitem [{\citenamefont {Meyer}\ and\ \citenamefont {Miller}(1979)}]{mey79}%
  \BibitemOpen
  \bibfield  {author} {\bibinfo {author} {\bibfnamefont {H.-D.}\ \bibnamefont
  {Meyer}}\ and\ \bibinfo {author} {\bibfnamefont {W.~H.}\ \bibnamefont
  {Miller}},\ }\href@noop {} {\bibfield  {journal} {\bibinfo  {journal} {J.
  Chem. Phys.}\ }\textbf {\bibinfo {volume} {70}},\ \bibinfo {pages} {3214}
  (\bibinfo {year} {1979})}\BibitemShut {NoStop}%
\bibitem [{\citenamefont {Stock}\ and\ \citenamefont {Thoss}(1997)}]{sto97}%
  \BibitemOpen
  \bibfield  {author} {\bibinfo {author} {\bibfnamefont {G.}~\bibnamefont
  {Stock}}\ and\ \bibinfo {author} {\bibfnamefont {M.}~\bibnamefont {Thoss}},\
  }\href@noop {} {\bibfield  {journal} {\bibinfo  {journal} {Phys. Rev. Lett.}\
  }\textbf {\bibinfo {volume} {78}},\ \bibinfo {pages} {578} (\bibinfo {year}
  {1997})}\BibitemShut {NoStop}%
\bibitem [{\citenamefont {Thoss}\ and\ \citenamefont {Stock}(1999)}]{tho99}%
  \BibitemOpen
  \bibfield  {author} {\bibinfo {author} {\bibfnamefont {M.}~\bibnamefont
  {Thoss}}\ and\ \bibinfo {author} {\bibfnamefont {G.}~\bibnamefont {Stock}},\
  }\href@noop {} {\bibfield  {journal} {\bibinfo  {journal} {Phys. Rev. A}\
  }\textbf {\bibinfo {volume} {59}},\ \bibinfo {pages} {64} (\bibinfo {year}
  {1999})}\BibitemShut {NoStop}%
\bibitem [{\citenamefont {{Schmidt am Busch}}\ \emph
  {et~al.}(2011)\citenamefont {{Schmidt am Busch}}, \citenamefont {M{\" u}h},
  \citenamefont {Madjet},\ and\ \citenamefont {Renger}}]{sch11}%
  \BibitemOpen
  \bibfield  {author} {\bibinfo {author} {\bibfnamefont {M.}~\bibnamefont
  {{Schmidt am Busch}}}, \bibinfo {author} {\bibfnamefont {F.}~\bibnamefont
  {M{\" u}h}}, \bibinfo {author} {\bibfnamefont {M.~E.-A.}\ \bibnamefont
  {Madjet}}, \ and\ \bibinfo {author} {\bibfnamefont {T.}~\bibnamefont
  {Renger}},\ }\href@noop {} {\bibfield  {journal} {\bibinfo  {journal} {J.
  Phys. Chem. Lett.}\ }\textbf {\bibinfo {volume} {2}},\ \bibinfo {pages} {93}
  (\bibinfo {year} {2011})}\BibitemShut {NoStop}%
\bibitem [{\citenamefont {Adolphs}\ and\ \citenamefont {Renger}(2006)}]{ado06}%
  \BibitemOpen
  \bibfield  {author} {\bibinfo {author} {\bibfnamefont {J.}~\bibnamefont
  {Adolphs}}\ and\ \bibinfo {author} {\bibfnamefont {T.}~\bibnamefont
  {Renger}},\ }\href@noop {} {\bibfield  {journal} {\bibinfo  {journal}
  {Biophys. J.}\ }\textbf {\bibinfo {volume} {91}},\ \bibinfo {pages} {2778}
  (\bibinfo {year} {2006})}\BibitemShut {NoStop}%
\bibitem [{Note1()}]{Note1}%
  \BibitemOpen
  \bibinfo {note} {T. C. Berkelbach and D. R. Reichman
  (unpublished)}\BibitemShut {NoStop}%
\bibitem [{\citenamefont {Jean}, \citenamefont {Friesner},\ and\ \citenamefont
  {Fleming}(1992)}]{jea92}%
  \BibitemOpen
  \bibfield  {author} {\bibinfo {author} {\bibfnamefont {J.~M.}\ \bibnamefont
  {Jean}}, \bibinfo {author} {\bibfnamefont {R.~A.}\ \bibnamefont {Friesner}},
  \ and\ \bibinfo {author} {\bibfnamefont {G.~R.}\ \bibnamefont {Fleming}},\
  }\href@noop {} {\bibfield  {journal} {\bibinfo  {journal} {J. Chem. Phys.}\
  }\textbf {\bibinfo {volume} {96}},\ \bibinfo {pages} {5827} (\bibinfo {year}
  {1992})}\BibitemShut {NoStop}%
\bibitem [{\citenamefont {Coalson}, \citenamefont {Evans},\ and\ \citenamefont
  {Nitzan}(1994)}]{coa94}%
  \BibitemOpen
  \bibfield  {author} {\bibinfo {author} {\bibfnamefont {R.~D.}\ \bibnamefont
  {Coalson}}, \bibinfo {author} {\bibfnamefont {D.~G.}\ \bibnamefont {Evans}},
  \ and\ \bibinfo {author} {\bibfnamefont {A.}~\bibnamefont {Nitzan}},\
  }\href@noop {} {\bibfield  {journal} {\bibinfo  {journal} {J. Chem. Phys.}\
  }\textbf {\bibinfo {volume} {101}},\ \bibinfo {pages} {436} (\bibinfo {year}
  {1994})}\BibitemShut {NoStop}%
\bibitem [{\citenamefont {Reid}\ \emph {et~al.}(1995)\citenamefont {Reid},
  \citenamefont {Silva}, \citenamefont {Barbara}, \citenamefont {Karki},\ and\
  \citenamefont {Hupp}}]{rei95}%
  \BibitemOpen
  \bibfield  {author} {\bibinfo {author} {\bibfnamefont {P.~J.}\ \bibnamefont
  {Reid}}, \bibinfo {author} {\bibfnamefont {C.}~\bibnamefont {Silva}},
  \bibinfo {author} {\bibfnamefont {P.~F.}\ \bibnamefont {Barbara}}, \bibinfo
  {author} {\bibfnamefont {L.}~\bibnamefont {Karki}}, \ and\ \bibinfo {author}
  {\bibfnamefont {J.~T.}\ \bibnamefont {Hupp}},\ }\href@noop {} {\bibfield
  {journal} {\bibinfo  {journal} {J. Phys. Chem.}\ }\textbf {\bibinfo {volume}
  {99}},\ \bibinfo {pages} {2609} (\bibinfo {year} {1995})}\BibitemShut
  {NoStop}%
\bibitem [{\citenamefont {Lucke}\ \emph {et~al.}(1997)\citenamefont {Lucke},
  \citenamefont {Mak}, \citenamefont {Egger}, \citenamefont {Ankerhold},
  \citenamefont {Stockburger},\ and\ \citenamefont {Grabert}}]{luc97}%
  \BibitemOpen
  \bibfield  {author} {\bibinfo {author} {\bibfnamefont {A.}~\bibnamefont
  {Lucke}}, \bibinfo {author} {\bibfnamefont {C.~H.}\ \bibnamefont {Mak}},
  \bibinfo {author} {\bibfnamefont {R.}~\bibnamefont {Egger}}, \bibinfo
  {author} {\bibfnamefont {J.}~\bibnamefont {Ankerhold}}, \bibinfo {author}
  {\bibfnamefont {J.}~\bibnamefont {Stockburger}}, \ and\ \bibinfo {author}
  {\bibfnamefont {H.}~\bibnamefont {Grabert}},\ }\href@noop {} {\bibfield
  {journal} {\bibinfo  {journal} {J. Chem. Phys.}\ }\textbf {\bibinfo {volume}
  {107}},\ \bibinfo {pages} {20} (\bibinfo {year} {1997})}\BibitemShut
  {NoStop}%
\bibitem [{\citenamefont {Neria}\ and\ \citenamefont {Nitzan}(1993)}]{ner93}%
  \BibitemOpen
  \bibfield  {author} {\bibinfo {author} {\bibfnamefont {E.}~\bibnamefont
  {Neria}}\ and\ \bibinfo {author} {\bibfnamefont {A.}~\bibnamefont {Nitzan}},\
  }\href@noop {} {\bibfield  {journal} {\bibinfo  {journal} {J. Chem. Phys.}\
  }\textbf {\bibinfo {volume} {99}},\ \bibinfo {pages} {1109} (\bibinfo {year}
  {1993})}\BibitemShut {NoStop}%
\bibitem [{\citenamefont {Shi}\ and\ \citenamefont {Geva}(2004)}]{shi04}%
  \BibitemOpen
  \bibfield  {author} {\bibinfo {author} {\bibfnamefont {Q.}~\bibnamefont
  {Shi}}\ and\ \bibinfo {author} {\bibfnamefont {E.}~\bibnamefont {Geva}},\
  }\href@noop {} {\bibfield  {journal} {\bibinfo  {journal} {J. Chem. Phys.}\
  }\textbf {\bibinfo {volume} {120}},\ \bibinfo {pages} {10647} (\bibinfo
  {year} {2004})}\BibitemShut {NoStop}%
\end{thebibliography}
%

\end{document}